\documentclass[epsf,twocolumn,showpacs,preprintnumbers]{revtex4}

\usepackage{amsmath,amssymb,amsfonts}
\usepackage{graphics}
\usepackage{dcolumn} 
\usepackage{bm,graphicx,color}
\usepackage{epsfig,comment}
\begin{document}
\title{Spin-Orbit Coupling Induced Degeneracy \\
  in the Anisotropic Unconventional Superconductor UTe$_2$}
\author{Alexander  B.  Shick}
\affiliation{Institute of Physics, Czech Academy of Science,  Na Slovance 2, CZ-18221 Prague, Czech Republic}
\author{Warren E. Pickett$^*$}
\affiliation{University of California Davis, Davis CA 95616} 
\email{pickett@physics.ucdavis.edu}
\date{\today}
\pacs{}

\begin{abstract}
The orthorhombic uranium dichalcogenide UTe$_2$ displays superconductivity
below 1.7 K, with the anomalous feature of retaining 50\% of normal state
(ungapped) carriers, according to heat capacity data from two groups. 
Incoherent transport that crosses over from
above 50 K toward a low temperature, Kondo lattice Fermi liquid regime 
indicates strong magnetic fluctuations
and the need to include correlation effects in theoretical modeling. We report
density functional theory plus Hubbard U (DFT+U) results for UTe$_2$ to provide 
a platform for modeling its unusual behavior, focusing on ferromagnetic (FM, time
reversal breaking) long range correlations along the ${\hat a}$ axis as established by
magnetization measurements and confirmed by our calculations.  
States near the Fermi level are
dominated by the $j=\frac{5}{2}$ configuration, 
with the $j_z=\pm\frac{1}{2}$ sectors being
effectively degenerate and half-filled. 
Unlike the small-gap insulating nonmagnetic electronic spectrum, 
the FM Fermi surfaces are large
(strongly metallic) and display low dimensional features, reminiscent of the
FM superconductor UGe$_2$. 
  
\end{abstract}
\maketitle

\section{Introduction}
Crystal symmetries have played a crucial role in classification of superconducting
gaps, with the well studied options being nodeless, point node, or line node gaps.
These classes have different types of low energy excitations, as observed
in spectroscopic, thermodynamic, and transport properties. Symmetries also
play a central role in the topological classification of normal crystalline materials,
leading to topological versus `trivial' insulators, Weyl semimetals, and 
multi-Weyl semimetals, as well as some more esoteric classes.
A conjunction of these criteria was discovered by Agterberg, Brydon, and Timm
(ABT)\cite{ABT2017}, who established the possibility of a superconducting phase
with topological protection that retains an {\it area} of gapless excitations, called
by them a Bogoliubov Fermi surface (BFS). The BFS phase combines a conventionally gapped
region of Fermi surface (FS) with an ungapped portion -- a normal electronic Fermi surface --
resulting in a new phase of matter which for now is called the BFS phase.

This theoretical development is especially prescient because of recent indications
of superconducting phases in which a finite fraction of carriers remain ungapped 
at temperature T=0. A definitive indication is a non-zero Sommerfeld specific heat
coefficient well below the superconducting critical temperature T$_c$, when the sample
is sufficiently free of second phases. One example is 
the Fe(S,Se) alloy system,\cite{Sato2018,Setty2019}
and other possible examples have been mentioned by ABT.\cite{ABT2017} The basic need at this
time is to obtain a realistic band structure including the requisite aspects that
will form the platform for inclusion of additional (viz. dynamic) effects.
We focus on the newly synthesized actinide chalcogenide UTe$_2$.

\begin{figure}[!htbp]
\centerline{\includegraphics[width=1.1\columnwidth]{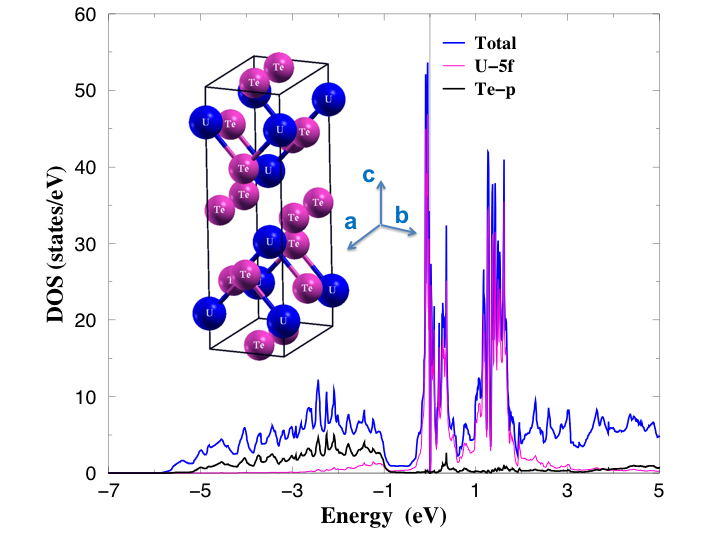}}
\caption{Total and projected densities of states/eV (per unit cell) for nonmagnetic
UTe$_2$ from the LSDA+U(OP) functional, with projections for U $5f$ (red) and
Te $p$ (black).
The $Immm$ UTe$_2$ crystal structure is shown in the inset: blue spheres are U,
pink spheres are Te. }
\label{dos}
\end{figure}

UTe$_2$, which crystallizes in an orthorhombic $Immm$ structure\cite{Ikeda2006} 
(space group \#71) shown in Fig.~\ref{dos}, has been suggested by Ran and 
collaborators\cite{Ran2018} to provide a new phase of superconducting matter 
below T$_c$=1.7K in which half of the electrons become superconducting and
half remain 
normal (thus with Fermi surfaces), based on heat capacity data.
The implication is that, at T$_c$, some additional symmetry is broken beyond
the usual broken gauge symmetry.
While no magnetic ordering is detected, it was proposed that half of the
fermionic excitations (spin up, for want of a better characterization) 
become superconducting 
while the spin down fermions remain in the normal conducting state. 
Confirmation of the basic properties has been provided by 
Aoki {\it et al.}~\cite{Aoki2019b},who extended some of the measurements. 
That this behavior occurs in a U-based compound brings to mind the three
U-based ferromagnetic superconductors UGe$_2$, URhGe, and UCoGe, for which
a recent review and comparison has been provided by Aoki, Ishida, and
Floquet.\cite{Aoki2019a}

Several features of UTe$_2$ have been revealed. The susceptibility is highly 
anisotropic, being much higher (as temperature is lowered) along the ${\hat a}$
axis. This anisotropy indicates that ferromagnetic ordering is more highly
favored along this axis, though ordering does not actually occur in zero
magnetic field. This anisotropy implies in turn a strong crystal 
field anisotropy of
the U atom, from which any magnetic moment must arise.   UTe$_2$ remains metallic
with a large (enhanced)  Sommerfeld coefficient $\gamma\approx 
120$ mJ/K$^2$,\cite{Ran2018,Aoki2019b} indicative of
a Kondo lattice fermionic ground state. Without any electronic structure study
for comparison, the degree of enhancement remains an open question. 
Aoki {\it et al.}\cite{Aoki2019b} have reported a 10\% entropy imbalance at T$_c$
between the observed state and the extrapolated normal state, which may
indicate some more complex behavior below T$_c$.

While U metallic $5f$ electrons become highly conducting at low T,
the magnetic susceptibility at higher temperature (above 150K) is characteristic
of a local Curie-Weiss moment, reported variously as somewhat anisotropic with 
values in the 3.3-3.6$\mu_B/U$ range,\cite{Ikeda2006} or also as 
2.8$\mu_B/U$.\cite{Ran2018} This local-itinerant
dichotomy is itself not so unusual, as
elemental Fe itself and several heavy fermion materials behave 
similarly. What is different about U is that 
spin-orbit coupling is very large, so pure spin differentiation gives way to
spin-orbit coupled quantum designations with the orbital contribution being
larger than that of the spin.  

The structure of UTe$_2$ is characterized by a shortest
U-U distance of 3.78\AA~ along a ``chain'' in the $a$ direction, with the next distance being
4.16\AA. This separation is well above Hill limit, implying that the $5f$ 
electrons should be localized and magnetic. 
The two U sites in the cell are related by inversion.
As mentioned, strong magnetic anisotropy in $\chi_{\alpha\alpha}(T)$ establishes 
${\hat a}$ as the easy axis, but this raises the question why  
no magnetic order down was detected down to 2K.

Calculations indicate that FM order is strongly favored over antiferromagnetic
ordering, when spins are aligned along the ${\hat a}$ direction. The MCA is very
large, nearly 100 meV, indicating that indeed FM alignment along ${\hat a}$ is 
favored, and order
must be being avoided by fluctuations. The large MCA indicates that 
crystal fields on the U site should be scrutinized.

We find, as expected, that U $5f$ states dominate the region around the 
Fermi level, described in Sec. III. Inclusion
of correlation effects appropriate for a more localized description versus a fully
itinerant description, by including a repulsive Hubbard $U$ interaction and a
Hund's $J$ parameter, do not change the primary features but do lead to an
effective degeneracy that may be relevant. The states at and
below E$_F$ are dominated by $j=\frac{5}{2}$ character (spin opposite to the
orbital moment). Within this multiplet, $m_j=-\frac{5}{2}$ and $-\frac{3}{2}$ 
states are fully occupied. The $m_j=\pm \frac{1}{2}$ states provide the
primary component of Fermi level states, and they are effectively degenerate
and half-filled. 
These projections are with respect to
the ${\hat a}$ axis, chosen as the majority spin direction. 
Furthermore, decomposition into spin states reveals that states around E$_F$
are essentially fully spin-polarized. These characteristics are suggestive of 
an incipient broken symmetry that might account for the Bogoliubov Fermi surface,
and in Sec.  IV some analysis is provided. A brief summary is provided in Sec. V.

\section{Methods}
The electronic structure has been investigated with the LSDA+U method,
reviewed and analyzed in Ref.~\cite{Ylvisaker2009}. 
Fully anisotropic repulsive orbital interactions $U_{m,m'}$ are included between all
U orbitals $m$, with Hund's coupling $J_{m,m'}$ between parallel spin electrons,
both treated in a fully rotationally invariant manner.\cite{SLP99}
In the case of strong spin-orbit coupling (SOC) appropriate for U,
to treat the strong orbital character realistically the spherical average
interaction  
\begin{eqnarray}
\Delta E^{ee}_{SS}=\frac{1}{2}({\bar U}-{\bar J})\Big(\mbox{Tr}[{\hat{n}}]
                       -\mbox{Tr}[{\hat{n}\hat{n}}] \Big) 
\end{eqnarray}
is considered as discussed in
Ref.~\cite{Dudarev1998}. Here ${\hat n}$ is the orbital occupation
matrix of the open shell and ${\bar U}, {\bar J}$ are spherically averaged
interactions. The remaining spin and orbitally dependent 
LSDA+U terms include the SOC induced anisotropic
contributions to the on-site Coulomb interactions -- the orbital 
polarization (OP) -- and the spin-flip terms due to spin off-diagonal 
matrix elements of the on-site occupation matrix $\hat{n}_{j_z,j_z'}$.

We use the relativistic version of the full-potential linearized augmented 
plane wave (FP-LAPW) method including SOC, with the 
rotationally invariant form of DFT+U implemented as described in Ref.~\cite{SLP99,SJDP2004}. 
An additional non-spherical double-counting correction is used, described
in Refs.~\cite{SJDP2004,Kristanovski18}. 
Following a conventional approach, we make 
use of reduced atomic Hartree-Fock values~\cite{Moore09} of the Slater 
integrals $F_2$ = 6.20 eV, $F_4$ = 4.03 eV, and $F_6$ = 2.94 eV. 
The resulting values are Hund's $J$ = 0.51 eV, and we select a 
Hubbard $U$ (=$F_0$) equal to the value of $J$. With the choice of the Coulomb repulsion $U$ equal to the Hund's exchange $J$,  
all spherically symmetric terms  
in the rotationally invariant $U, J$ correction are set to zero,
as they are treated in the LSDA functional.
This approach 
can be regarded as the orbital polarization (OP) limit of LSDA+U;
orbital polarization functionals with DFT/\cite{OPC,Eschrig} might
also be tried. 
For complete clarity,
the functional we use is presented in detail in Appendix A.

\section{Analysis of Results}
\subsection{Non-magnetic LDA+U(OP)}
 As a reference point we first perform non-magnetic LSDA+U(OP) calculations. 
The electronic density of states (DOS), together with the U atom 
$f$-projected DOS and Te $p$-projected DOS, are shown in Fig.~\ref{dos}. 
The $\frac{5}{2}$ and $\frac{7}{2}$ peak centers are separated by $\sim$1 eV.
The $\xi \vec s\cdot\vec\ell$ term in the Kohn-Sham equation has a coefficient
$\xi_{5f}$=220 meV, giving a $\frac{5}{2}-\frac{7}{2}$ splitting of 0.77 eV,
confirming that this separation is from SOC, with minor crystal field
contributions.
The calculated 5$f$ occupation within the U sphere $n_f$= 2.8, 
which supports the viewpoint of an underlying $f^3$ configuration. Additional
evidence for this will appear later. A filled Te $p$ shell, {\it i.e.} Te$^{2-}$,
would require U$^{4+}$ $f^2$, so the Te $p$ shell is not filled and,
in spite of the small U 5$f$ bandwidth, U-Te hybridization cannot be discounted.

The narrow 5$f$ electron states shown in Fig.~\ref{dos} that 
straddle the Fermi energy ($E_F$) in Fig.~1 have $j=\frac{5}{2}$ character; 
the $j=\frac{7}{2}$ manifold is centered
1.5 eV above $E_F$. Careful determination of the bands and DOS reveals a
gap of 13 meV, apparently an ``accidental'' hybridization gap rather than one
between characteristic bands (bonding-antibonding, specific $m_j$ characters,
etc.)
Due to a flatness of the $f$ bands there is a strong peak in the DOS (up to 35
states/eV) just 15 meV below $E_F$ which would correspond to
 $\gamma$= 41 mJ K$^{-2}$ mol$^{-1}$. Even that value is much smaller than
the experimental value of $\gamma$=120 mJ K$^{-2}$ mol$^{-1}$~\cite{Ran2018}, 
indicating a substantial dynamical enhancement which, considering the Kondo-like
behavior observed in the resistivity, is likely due to magnetic fluctuations.

\begin{figure}[!htbp]
\centerline{\includegraphics[width=1.0\columnwidth]{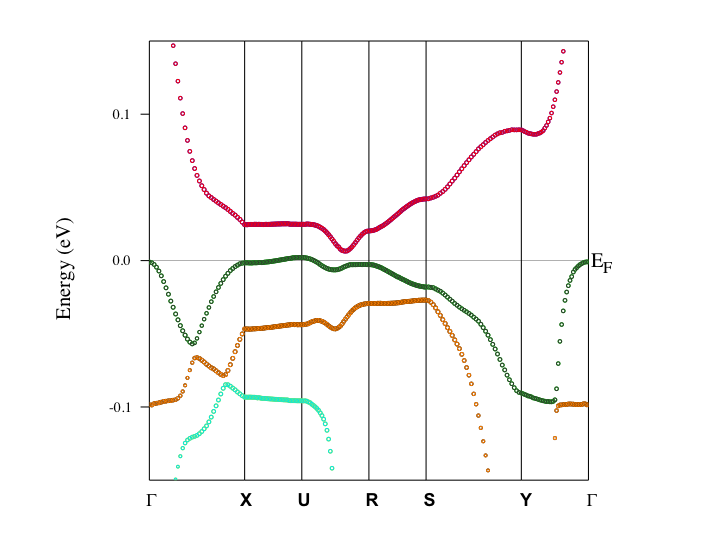}}
\caption{The band structure and the  FS from non-magnetic
LSDA+U(OP) calculations. The circle size indicates the amount of $f$-character of the
eigenstates. 
}
\label{bnd}
\end{figure}

An enlargement of the band structure near $E_F$ is shown in Fig.~\ref{bnd}. 
Simple non-magnetic, uncorrelated UTe$_2$ is calculated to be a semimetal, 
similar to that found by Aoki {\it et al.}~\cite{Aoki2019b} using LDA 
without correlation corrections.  The 13 meV bandgap 
just above $E_F$ reflects a hybridization gap occurring between a heavy and a
 light band, although the band structure is more involved than the textbook
viewpoint. The semimetallic character of nonmagnetic UTe$_2$ might suggest
an instability toward electron-hole pairing~\cite{Volkov75} 
though there is no evidence of such a new phase. Also, we find in the next
section that OP changes the Fermi level electronic structure substantially.

The $|m_l,m_s >$ and $|m_j=m_l+m_s >$ decompositions of $N(E_F)$ are provided
in Table~\ref{dosef} of Appendix B.
Since the experimental susceptibility is highly anisotropic at low T, 
being much higher along the $\hat{a}$ axis, we chose it as the 
moment quantization axis
and it will be used below in the results for ferromagnetic alignment.
The major contribution to the FS comes from $m_l=0, m_s=-\frac{1}{2}$ and
$m_l=-1, m_s=+\frac{1}{2}$ in equal amounts, providing the orbital and
spin composition of the active orbitals. 
Given the strong SOC of U,
the angular momentum coupled representation is more fundamental.  
The decompositions    
  $|j=\frac{5}{2},m_j=\pm\frac{1}{2}>$ states (henceforward,
this notation will be $|\frac{5}{2},\pm\frac{1}{2}>)$
 plus some non-negligible amount of  $|\frac{5}{2},\pm\frac{5}{2}>$ states
 are dominant.

In Appendix C information on the composition of the bands near E$_F$ is provided.
The ``fat-band" structure  is shown in Fig.~\ref{bnd2} emphasizing the 
 $|m_l,m_s >$ and $|m_j=m_l+m_s>$ character of the $j=\frac{5}{2}$ manifold,
for negative $m_j$.

\subsection{Ferromagnetic LSDA+U(OP)}
No magnetic order has been detected down to 0.4K. However, based on the 
observed susceptibility and strong magnetocrystalline anisotropy (MCA)
from both theory and experiment, slow long-range FM correlations with
alignment along the ${\hat a}$-axis are expected, {\it i.e.} locally the
electronic structure is FM. 
To model this low T phase we have performed FM calculations with spin 
moments (and in this collinear calculation, the orbital moments as well)
 aligned along the $\hat{a}$ axis.
The FM state is 185 meV/f.u. lower in the energy than the non-magnetic state.
The $f$-shell ordered moments, oriented along the ${\hat a}$-axis, 
are calculated to be \\~~~
$\vec M\parallel {\hat a}$: $M_S$=1.92$\mu_B$, $M_L$=-3.44$\mu_B$, $M_J$=-1.52$\mu_B$,\\
$\vec M\parallel {\hat b}$: $M_S$=1.83$\mu_B$, $M_L$=-3.71$\mu_B$, $M_J$=-1.88$\mu_B$,\\
$\vec M\parallel {\hat c}$: $M_S$=1.91$\mu_B$, $M_L$=-3.97$\mu_B$, $M_J$=-2.06$\mu_B$.\\
An analogous calculation for FM UGe$_2$ gave 
$M_S$=1.32$\mu_B$, $M_L$=-2.92$\mu_B$, $M_J$=-1.58$\mu_B$, 
which was in good agreement with experimental data~\cite{Kernavanois2001} in
the ordered state.

The Curie-Weiss moment, an average over fluctuations in all dimensions,
is 2.8$\mu_B$ [\onlinecite{Ran2018}] to 3.3$\mu_B$ [\onlinecite{Aoki2019b}] for UTe$_2$. 
It is common, when moments are not strongly localized, that the ordered moment
is reduced due to mixing with the conduction states.  Lack of ordering in UTe$_2$
precludes comparison, but substantial itinerant character of the moment is apparent.
Calculating the magnetocrystalline anisotropy, we find that for moments 
oriented along the $\hat{b}$ and $\hat{c}$ axes,
the energy is higher by 37 meV/f.u. and 96 meV/f.u. respectively. Thus,
there is a strong uniaxial magnetic anisotropy, with $\hat{a}$ being the easy axis.

\begin{figure}[!htbp]
\centerline{\includegraphics[width=1.0\columnwidth]{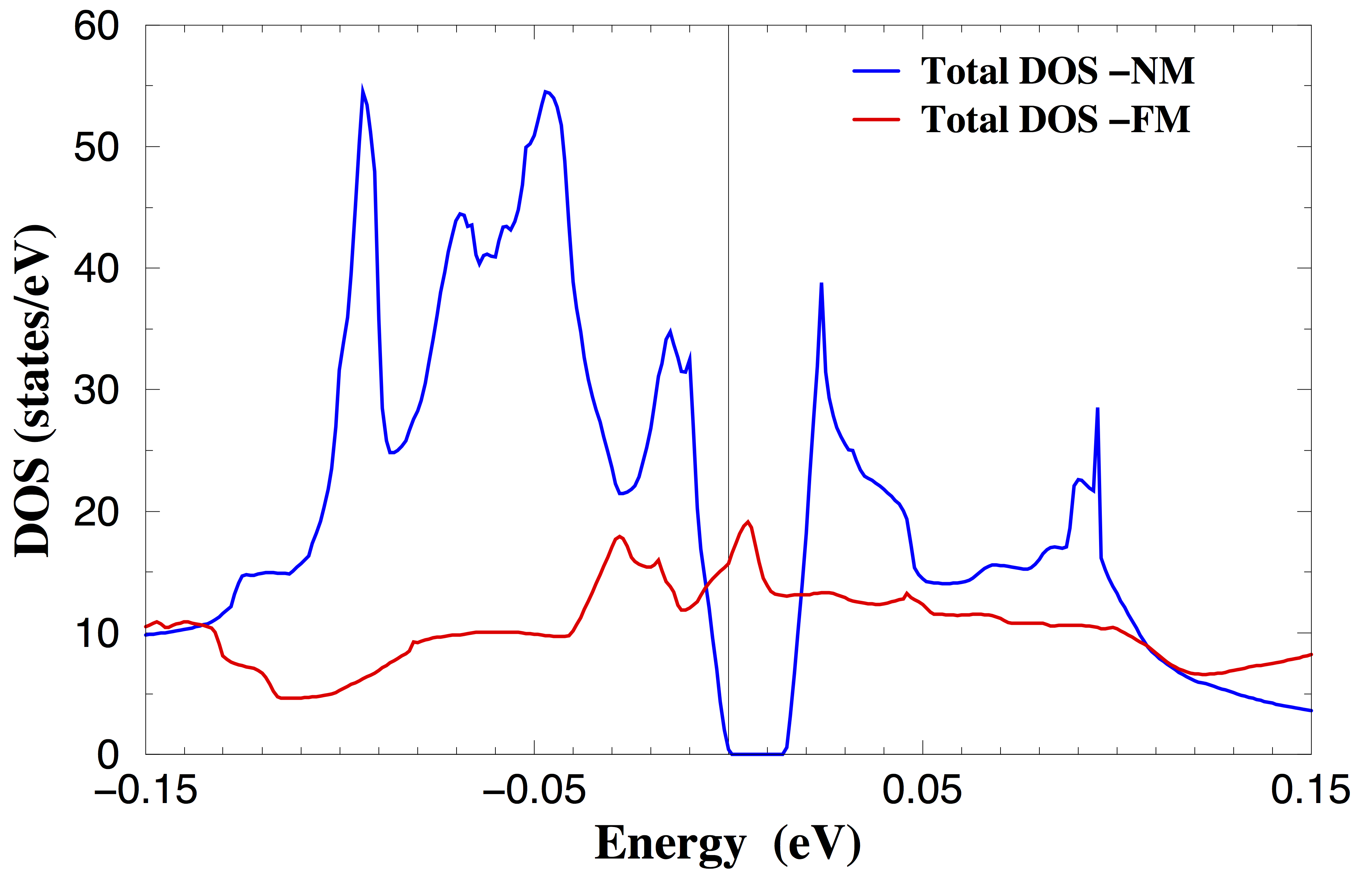}}
\caption{Comparison of the density of states for ferromagnetic alignment
(metallic) with that of the gapped nonmagnetic state, on a fine energy scale. 
The 13 meV gap is thoroughly washed out when magnetic moments are
allowed to emerge and align.
}
\label{2dos}
\end{figure}

The effect of allowing FM moment development and alignment is displayed in
Fig.~\ref{2dos}. The strong spin- and orbital-polarization coupled with strong
SOC results in washing out of the small hybridization gap calculated for the nonmagnetic
system. The value of N($E_F$) is 16 states/eV-cell corresponding to a band
value of $\gamma_{\circ}$= 19 mJ/K$^2$, leaving a factor of five enhancement due
to dynamic processes to account for the observed Sommerfeld coefficient. 

\begin{figure}[!htbp]
\centerline{\includegraphics[width=1.0\columnwidth]{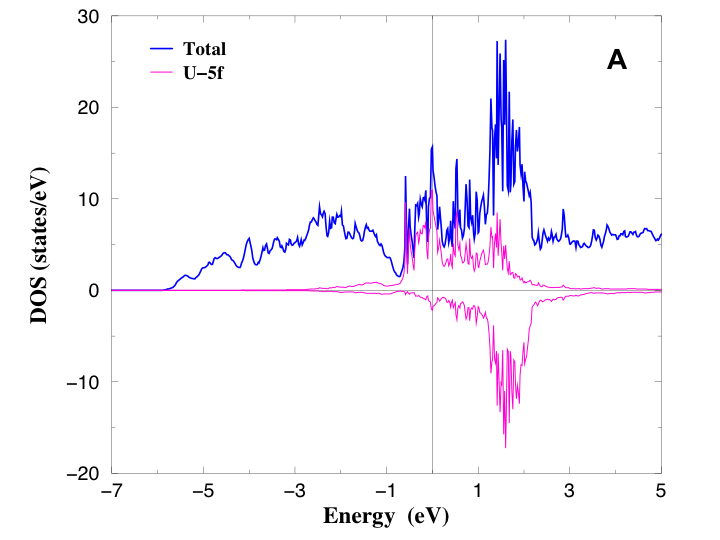}}
\centerline{\includegraphics[width=1.0\columnwidth]{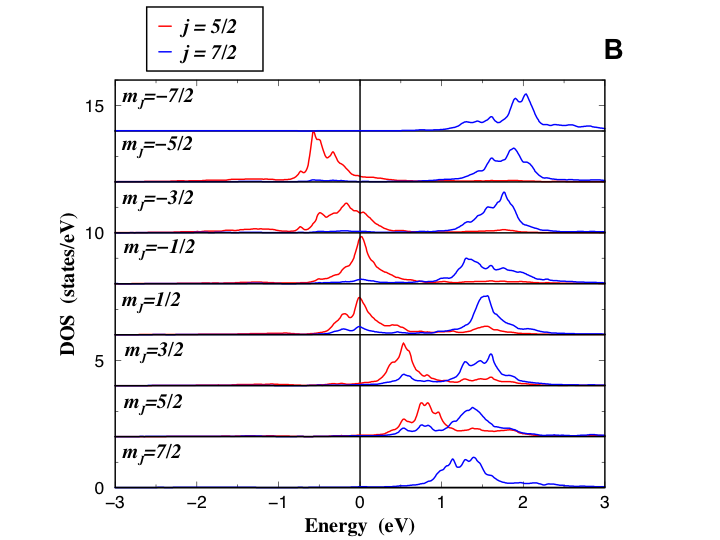}}
\caption{
Total (A) and $f$-projected (B) densities of states for ferromagnetic
UTe$_2$ from LSDA+U(OP) functional, with the moment along ${\hat a}$. (A)
The $j_z$-decomposed DOS for both $j=\frac{5}{2}$, plotted upward in red, 
and $j=\frac{7}{2}$, plotted downward. (B)
The $j_z$ projected density of states, with projections stated in the panels. 
Occupied states arise almost entirely from $j=\frac{5}{2}$
states with negative $j_z$.
}
\label{dosFM}
\end{figure}

The density of states for FM ordered UTe$_2$, projected onto spin directions,
is displayed in Fig.~\ref{dosFM}A. The exchange splitting is seen to be around 1.5 eV,
with negligible population of minority spin 5$f$ orbitals -- the $f$ shell of U
is fully spin-polarized, however SOC mixes the spin moment amongst 
the $m_j$ states. The U 
$j_z$-projected DOS for both $j=\frac{5}{2}$ and $j=\frac{7}{2}$ subspaces
is pictured in Fig.~\ref{dosFM}B. The fully occupied states arise  
from $j=\frac{5}{2}$ states with $j_z=-\frac{5}{2}$ and $-\frac{3}{2}$. The
$j_z=\pm\frac{1}{2}$ states are half-filled, a point we return to below;
higher $j_z$ states are unfilled. The U atom configuration can thus be
characterized as an $f^2$ local moment, with $j_z=\pm\frac{1}{2}$ orbitals
that are itinerant and whose spins compensate.

\begin{figure}[!htbp]
\centerline{\includegraphics[width=1.0\columnwidth]{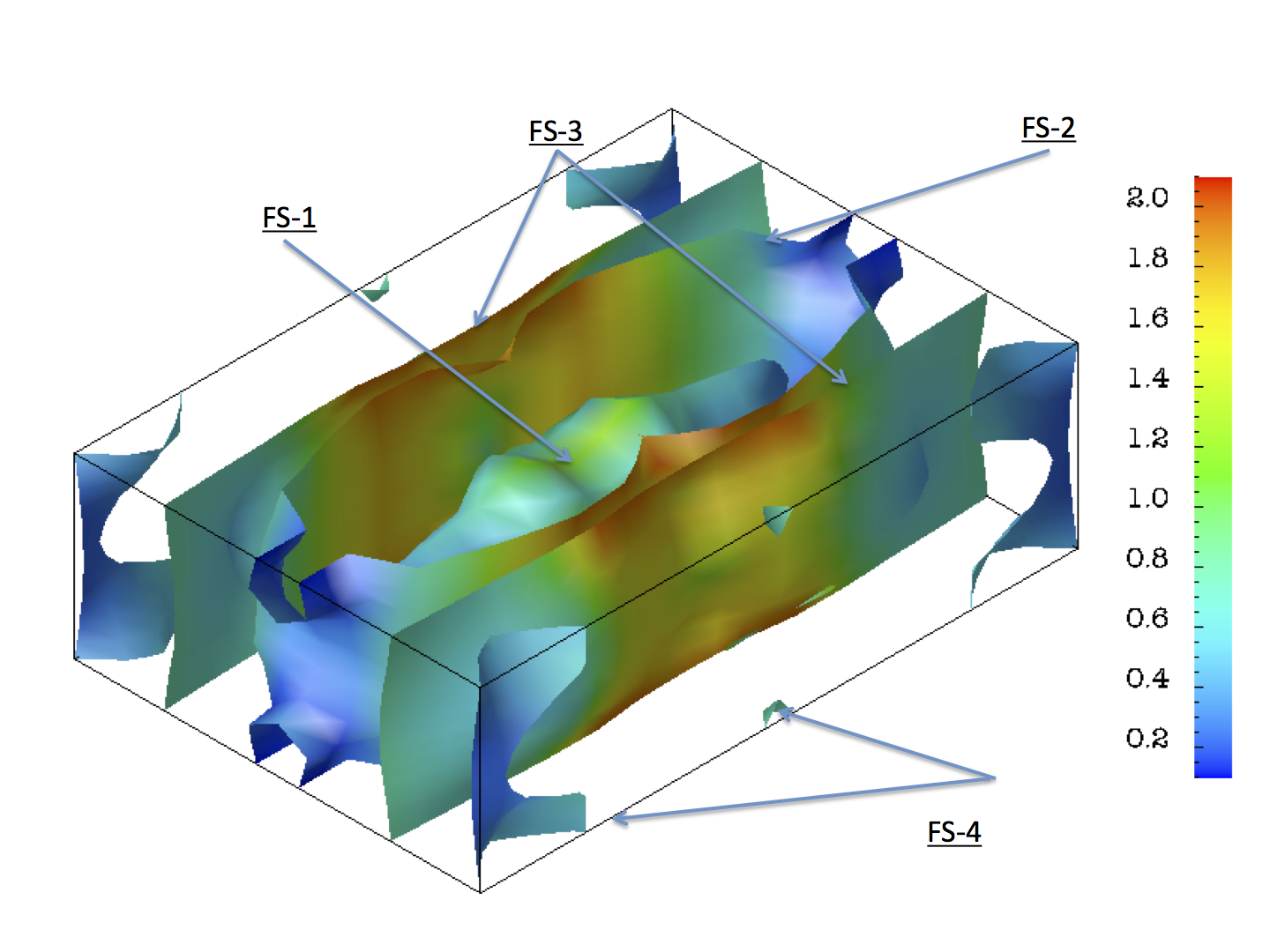}}
\caption{
Fermi surface for FM order, from LSDA+U(OP),
centered at the $\Gamma$ point.
The Fermi surface is large and multi-sheeted, with  a
nesting feature for FS-3 near (0, ${\frac{\pi}{b}}$, 0).
FS-1 is in the center; FS-2 parallels FS-3 along parts of its length.
The low velocities, given by the colorbar,  extend over at least an order
of magnitude, up to nearly 2 eV-\AA ($\sim 2.5\times 10^7$ cm/s).
}
\label{sp-fs}
\end{figure}

The Fermi surface for FM order is displayed in Fig.~\ref{sp-fs}.
It has 4 sheets, with the most obvious characteristic being a 
strong nesting feature for FS-3 near 
(0, ${\frac{\pi}{b}}$, 0). Such nearly parallel sheets, one-dimensional in
character, suggest instabilities toward order that would
double the cell along the ${\hat b}$ axis.
The $|m_j=m_l+m_s>$ decompositions of $N(E_F)$ are given in Table~\ref{dosef2} of 
Appendix B.
The corresponding $|m_l,m_s>$ decompositions of $N(E_F)$ are provided in 
Appendix B in Table~\ref{dosef3}. From Table~\ref{dosef2}, it can be
observed that FS-2 has strong $j_z=\frac{1}{2}$ character while FS-4 has
strong $j_z=-\frac{1}{2}$ character, with some $j_z=\pm\frac{1}{2}$  character
spread over the other sheets.

\section{Considerations for Partial Pairing}
Heat capacity data\cite{Ran2018,Aoki2019b} indicates that half of the
fermionic carriers remain ungapped
below T$_c$, the other half gapped by superconductivity. A central issue is
that no information is yet available on the ``halves'' that might be
involved.  With a multisheeted FS with complex shape,
it is difficult to identify half of the Fermi surface for
gapping, barring the trivial and unphysical dividing of the zone into
halves, quarters, or eights according to the $Immm$ symmetry. The 
spin-up -- spin-down possibility viewpoint,
for example, does not have an obvious proclivity to gap exactly half of a
complex Fermi surface when SOC is strong, while pairing only one spin
would suggest time-reversal symmetry breaking.

We have suggested that at low T near T$_c$, the long-range and slow magnetic
fluctuations imply that  UTe$_2$ is locally FM. In this
case spin degeneracy is broken, and moreover strong SOC mixes the two spins
with orbital characters. The bands and Fermi surfaces are non-degenerate, but 
inversion provides $\varepsilon_{-k}=\varepsilon_k$ as the only degeneracy.
Averaging over the moment fluctuations restores moment symmetry on a time
scale slower than that of the fluctuation time. The next consideration is that
the resistivity indicates that Kondo coherence has restored a translationally
invariant state (``screened the moments'') that does not scatter carriers. 
This coherence and lack of broken symmetry suggests that degeneracy, viz.
up-down, is restored. On the other hand, the large susceptibility implies that strong
moment fluctuations remain present.

\subsection{Local spin-orbital considerations.}
The degeneracy of the $m_j=\pm\frac{1}{2}$ orbitals and their half filling 
suggests considering linear combinations of the degenerate orbitals, 
which are themselves
linear combinations of $5f$ orbitals and spin projections. The
spin-orbit coupled combinations of $m_j=\pm\frac{1}{2}$ states 
are\cite{landaulifshitz}
\begin{eqnarray}
\phi_{\frac{5}{2},+\frac{1}{2}}&=&-\sqrt{\frac{3}{7}}Y_{3,0}|\uparrow>
                                 +\sqrt{\frac{4}{7}}Y_{3,+1}|\downarrow> \nonumber \\
\phi_{\frac{5}{2},-\frac{1}{2}}&=&-\sqrt{\frac{3}{7}}Y_{3,-1}|\uparrow>
                                 +\sqrt{\frac{4}{7}}Y_{3,0}|\downarrow>
\end{eqnarray}
in terms of spherical harmonics $Y_{\ell,m}$ and spin projections. Each state
contains $Y_{3,0}$ but with opposite spin projections.  Rotating
these degenerate orbitals in functional space so the common orbital $Y_{3,0}$
has equal spin-up and -down amplitudes leads to 
\begin{eqnarray}
\Phi_{\pm}&=&
       [\sqrt{\frac{4}{7}}\phi_{\frac{5}{2},+\frac{1}{2}}
\pm \beta   \sqrt{\frac{3}{7}}\phi_{\frac{5}{2},-\frac{1}{2}}],
        \nonumber \\
 &=& [\sqrt{\frac{16}{49}} Y_{3,+1}|\downarrow>
 \mp \beta\sqrt{\frac{ 9}{49}} Y_{3,-1}|\uparrow>] \nonumber \\
 & &+  \sqrt{\frac{24}{49}} Y_{3,0}\frac{[-|\uparrow>\pm \beta|\downarrow>]}{\sqrt{2}}
\end{eqnarray}
where $\beta=1$ or $\beta=i$.
This linear combination leads to the 
orbital function $Y_{3,0}$ with vanishing orbital moment and a specific
 spin projection: $<\vec s> = (0,\pm 1,0$, which has zero projection along
the orbital $z$-axis, which due to the large magnetic anisotropy is the
crystalline ${\hat a}$-axis. 
The other part has entwined ($Y_{3,\pm 1}$) orbitals with opposite 
spin projections. 
A second intriguing observation is that one part has a 
fraction $\frac{25}{49}$=51\% of the weight, while
the second contains 49\%; these fractions are experimentally indistinguishable
from the ``one half'' of carriers that become paired at T$_c$ in UTe$_2$.
These observations suggest possible symmetry breaking that links the magnetic
anisotropy with the orbital occupation, a possible line for further consideration.

\subsection{The coherent-carrier itinerant picture}
The Fermi liquid behavior, evidenced by the strong decrease in the resistivity by a
factor of up to 35 below 50K,\cite{Aoki2019b} reflects Kondo screening of the $5f$ moments.
While the large susceptibility, arising from field alignment of magnetic
fluctuations, indicates dynamic moments, they no longer scatter carriers,
primarily just contributing a mass enhancement which we can estimate is of the
order of a factor of five. 

Aoki, Ishida, and Floquet have observed in their review\cite{Aoki2019a} that only
uranium compounds have been confirmed as FM superconductors: orthorhombic U$X$Ge,
$X$=Ge, Rh, Co. The latter two are isovalent.  We obtain an $f^3$ configuration 
in UTe$_2$, just as has been concluded experimentally for UCoGe,\cite{Aoki2019a}
so the $f^3$ configuration may carry importance.

For weak ferromagnets and incipient ferromagnets, DFT methods are known to
overestimate the tendency toward magnetic order and the size of magnetic moments, 
usually ascribed to neglect of
magnetic fluctuation effects in the functionals, and possibly to
the necessity to include dynamic effects explicitly. Our work provides the
groundwork for modeling the metallic Fermi liquid phase of UTe$_2$ that provides
the platform for an exotic superconducting state. In UPt$_3$ for example, the
six-sheeted Fermi surface is given very realistically by 
LDA methods.\cite{upt3_nrl1,upt3_nrl2,upt3_nrl3,upt3_nrl4,mcmullan} In several
cases, however, Fermi surfaces are not predicted well by first principles methods.

Our work provides important guidelines: 
the heavy fermion Fermi liquid state of UTe$_2$ is built on an $f^3$ U
configuration in which $m_j=\pm \frac{1}{2}$ states in the $j=\frac{5}{2}$ 
subspace provides the primary orbital content of the Fermi surface states
$|k,n,\tau>$. Here $\tau$ indicates the pseudospin component that bears some
characteristics of the more common spin degree of freedom that is commonly used
to categorize exotic pairing states.  The specific spin-orbital 
decompositions discussed in the
previous subsection may deserve further consideration.

\section{Discussion and Summary}
UTe$_2$ presents the usual challenges of a Kondo lattice superconductor, with the complicating
features of strong magnetic anisotropy (strong SOC) and (especially) 
the coexistence of normal carriers
and superconducting pairs below T$_c$.   
At low T magnetic fluctuations slow down, especially those with large moments
as observed in UTe$_2$. Considering the strong evidence, both experimental and
theoretical, of an easy $\hat a$
axis and strong magnetocrystalline anisotropy, 
a outstanding question is how UTe$_2$ manages to avoid magnetic order. Kondo screening
of the moments is the primary rationalization.

Comparison and contrast of UTe$_2$ and UPd$_2$Al$_3$ 
[\onlinecite{upd2al3}] are striking. Both have 
comparable specific heat $\gamma$'s indicating Kondo lattice character
arising from spin fluctuations, and both become superconducting at 1.7-1.8K.
Both can be characterized as a local moment from a localized $f^2$ 
pair or orbitals, with roughly one more $5f$ electron being itinerant.
However, UPd$_2$Al$_3$ orders antiferromagnetically at six times higher than
T$_c$, compared to UTe$_2$ which (according to muon spin rotation 
data\cite{muSR}) does not order magnetically down to 25 mK. And of course,
only half the carriers in UTe$_2$ become superconducting. Evidently the
phases in these intermetallic uranium compounds are sensitive to several
relative energy scales.

We have explored the possibility that at low T (somewhat above the superconducting T$_c$) 
there is medium range FM
orientation of the U moments, from $j=\frac{5}{2}$ states (quantization axis is the
$\hat a$ axis) which have strong localized
character (moments around 2-3$\mu_B$) but strong enough hopping to enable conducting
behavior with correlation enhancements. In this case the U sites are locally 
ferromagnetic, involving essentially pure
spin (majority) moments. Magnetic (exchange) coupling proceeds through the spin
moments, so the single spin character of the calculated moments become relevant.
In the limit of negligible SOC and one spin channel being frozen out,
triplet pairing reduces to single-spin superconducting pairing\cite{sss1,sss2}
suggested as a possibility in half-metals.

As mentioned in the Introduction, Agterberg and collaborators have pointed out
the possibility of symmetry restricted ``Bogoliubov Fermi surfaces:'' 
areas of Fermi surfaces that remain
ungapped in the superconducting state and give, among other effects, a non-zero
Sommerfeld coefficient above the superconducting ground state. With no obvious
symmetry(s) beyond the space group in the calculated Fermi surface and global
time-reversal symmetry, there is no apparent
reason why the normal fraction should be (within uncertainty) half of the number
of carriers in the normal state. Can this ``50\%'' fraction be accidental? It
hardly seems likely. 

We have presented a scenario for this 
division: the Clebsch-Gordon coupling enforced by strong SOC suggests a coupling
that favors the choice of spatially uniform, orbital-moment free, spin-mixed
spin orbitals as the building blocks for the Bloch orbitals that will eventually
pair. The linear combination that suggests the importance of these orbitals
contains 49\% of the Fermi level spectral density of $j_z=\pm \frac{1}{2}$ states
that are calculated to be effectively degenerate in the correlated band structure of
this heavy fermion compound.  The half-filled nature 
of these $j_z=\pm \frac{1}{2}$ orbitals, with density of states peaking at E$_F$,
suggests the tendency toward symmetry-breaking states.

The simplest scenario for the apparent Bogoliubov Fermi surface carriers observed
in superconducting UTe$_2$ is the simple separation (spin-up versus -down) 
used by Ran {it et al.}, adapted to this compound. In terms of pairing of the
$j_z=\pm\frac{1}{2}$ orbitals, triplet pairing involves the space of $J=0, \pm 1$
pairs
\begin{eqnarray}
\Delta(k) = d_0 \Delta_0(k) + d_1 \Delta_1(k) + d_{-1} \Delta_{-1}(k)
\end{eqnarray}
in the usual notation.\cite{mineev2017} 
Strong on-site ferromagnetic alignment raises the first 
terms with its antialigned moments to high energy. Then the vanishing of
(say) $\Delta_{-1}$ results in a superconducting phase with $J=1$ pairs gapped
and the remaining 50\% of carriers residing on a Bogoliubov Fermi surface.
The microscopic source of such symmetry breaking, if it occurs, remains for
further studies.

\section{Acknowledgments}
W.E.P. thanks P. Hirschfeld for an introduction to the Bogoliubov Fermi surface
literature and concepts, and comments on the manuscript.
A.B.S. acknowledges partial support from MSMT Project 
No. SOLID21-CZ.02.1.01/0.0/0.0/16$_{-}$019/0000760,
and GACR grant No. 18-02344S. W.E.P. was supported by NSF Grant DMR 1607139.

\appendix
\section{The exchange-correlation functional}
The electron-electron interaction energy
$E^{ee}$ in the DFT+U total-energy functional, with the
spin-orbit coupling (SOC) included \cite{igor}, has the form
\begin{eqnarray}
\label{eq:1} E^{ee} = \frac{1}{2}  \sum_{\bf \gamma_1 \gamma_2
\gamma_3 \gamma_4} n_{\bf \gamma_1 \gamma_2} \Big( V^{ee}_{\bf
\gamma_1 \gamma_3; \gamma_2 \gamma_4} - V^{ee}_{\bf \gamma_1
\gamma_3;\gamma_4 \gamma_2} \Big) n_{\bf \gamma_3 \gamma_4} \; ,
\end{eqnarray}
which contains the 14x14 on-site occupation matrix $n_{\bf \gamma_1
\gamma_2} \equiv n_{m_1 \sigma_1,m_2 \sigma_2}$ with 
generally non-zero orbital and spin off-diagonal matrix elements.
The $V^{ee}$ is an effective on-site Coulomb interaction,
expressed in terms of Slater integrals (see  Eq.(3) in Ref.~[\onlinecite{SLP99}]) 
which are linked to 
intra-atomic repulsion $U_{m,m'}$ and
exchange $J_{m,m'}$ quantities mentioned in Sec. II. The spherically symmetric 
double-counting energy $E^{dc}$ is subtracted from  $E^{ee}$ to correct on 
the electron-electron interaction already included in DFT.

The DFT+U energy correction $\Delta E^{ee} =  E^{ee} - E^{dc}$ can be divided 
into a sum of spherically symmetric and anisotropic terms. 
In the case of  the "atomic" or "fully localized" (FLL)  limit of $E^{dc}$ , and without SOC,
the spherically-symmetric part is given by~\cite{Dudarev1998},   
\begin{eqnarray}
\label{eq:2} \Delta E^{ee} = \frac{(U-J)}{2} \Big( \mbox{Tr}[{\hat{n}}] - \mbox{Tr}[{\hat{n} \hat{n}}] \Big) 
\end{eqnarray}

The choice  $U = J$ in Eq.~(\ref{eq:1}) means that the spherically symmetric part of $\Delta E^{ee}$
given by Eq.~(\ref{eq:2}) becomes equal to zero. The remaining non-spherically symmetric part of  $\Delta E^{ee}$
can be regarded as the DFT+U analog of the proposed
"orbital polarization correction" functionals (OPC).\cite{OPC,Eschrig}  

Due to the full potential character, care should be taken to exclude the
so-called "non-spherical double counting" of the $f$-state nonspherical contributions to the DFT and DFT+U 
parts of the Kohn-Sham potential. 
When the atomic sphere matrix elements of the DFT+U Hamiltonian are calculated, 
those contributions from the lattice harmonics $K_{\nu}$ of the nonspherical part of the DFT potential
$V_{DFT}^{NSH}({\bf r}) = \sum_{\nu} V_{\nu}(r) K_{\nu}(\hat{r})$ are removed, 
which are proportional to
$\langle lm_1 | K_{\nu} | lm_2 \rangle$ for $l=3$, the $f$-states orbital quantum number.

\section{Decomposition of the U $5f$ DOS at $E_F$}

This table provides the stated decomposition of the Fermi level density
of states.

\begin{table}[!htbp]
\caption{The $|m_l, m_s >$ and $|j,m_j>$ decompositions of
the U atom $f$-projected DOS at $E_F$ (in states/eV unit cell)
for the unpolarized system.
The magnetic quantization is along the easy $\hat{a}$ axis.}
\label{dosef}
\begin{ruledtabular}
\begin{tabular}{ccccccccc}
$m_l$  & -3  & -2 & -1 & 0 & 1 & 2 & 3 \\
\hline
spin-$\uparrow$     & 0.16 & 0.04 & 0.10 & 0.16 & 0.01 & 0.03 & 0.00 \\
spin-$\downarrow$& 0.00 & 0.03 & 0.01 & 0.16 & 0.10 & 0.04 & 0.16 \\
\hline
$m_j$  & -7/2  & -5/2 & -3/2 & -1/2 & 1/2 & 3/2 & 5/2 & 7/2 \\
            & 0      & 0.19 & 0.05 & 0.26 & 0.26 & 0.05 & 0.19 & 0 \\
\end{tabular}
\end{ruledtabular}
\end{table}

\begin{table}[htbp]
 \caption{Decomposition of the four Fermi surface contributions to $N(E_F)$
from $|m_j=m_l+m_s>$ projections, for FM aligned UTe$_2$.
Magnetic quantization is along the $\hat{a}$ axis. The
dominance of the $j_z=\pm\frac{1}{2}$ is evident.}
\label{dosef2}
\begin{ruledtabular}
\begin{tabular}{ccccccccc}
$m_j$  & -7/2   & -5/2 & -3/2 & -1/2 & 1/2  & 3/2  & 5/2  & 7/2 \\
\hline
FS1    & 0      & 0.02 & 0.17 & 0.15 & 0.25 & 0.01 & 0.03 & 0.01 \\
FS2    & 0      & 0.08 & 0.21 & 0.71 & 1.35 & 0.06 & 0.03 & 0.01 \\
FS3    & 0      & 0.05 & 0.23 & 0.63 & 0.36 & 0.04 & 0.02 & 0.01 \\
FS4    & 0      & 0.04 & 0.27 & 1.30 & 0.48 & 0.04 & 0.02 & 0.01 \\
\hline
Sum    & 0      & 0.19 & 0.88 & 2.79 & 2.44 & 0.15 & 0.10 & 0.04 \\
\end{tabular}
\end{ruledtabular}
\end{table}

\begin{table}[htbp]
\caption{The $|m_l,m_s>$  decomposition of $N{_f}(E_F)$ (eV$^{-1}$), for 
ferromagnetic alignment of UTe$_2$ along the ${\hat a}$ axis. The 
down spin components are small; as noted in the text, the system is
near full spin polarization..
}
\label{dosef3}
\begin{ruledtabular}
\begin{tabular}{ccccccccc}
  \multicolumn{8}{c}{FS1} \\
$m_l$  & -3  & -2 & -1 & 0 & 1 & 2 & 3 \\
\hline
spin-$\uparrow$     & 0.02 & 0.14 & 0.10 & 0.20 & 0.01 & 0.02 & 0.01 \\
spin-$\downarrow$& 0.00 & 0.00 & 0.03 & 0.05 & 0.05 & 0.00 & 0.01 \\
\hline
 \multicolumn{8}{c}{FS2} \\
\hline
spin-$\uparrow$     & 0.08 & 0.18 & 0.60 & 1.12 & 0.04 & 0.02 & 0.01 \\
spin-$\downarrow$& 0.00 & 0.00 & 0.03 & 0.11 & 0.23 & 0.02 & 0.01 \\
 \hline
 \multicolumn{8}{c}{FS3} \\
\hline
spin-$\uparrow$     & 0.05 & 0.20 & 0.52 & 0.30 & 0.02 & 0.02 & 0.01 \\
spin-$\downarrow$& 0.00 & 0.00 & 0.03 & 0.11 & 0.06 & 0.02 & 0.01 \\
\hline
 \multicolumn{8}{c}{FS4} \\
spin-$\uparrow$     & 0.04 & 0.24 & 1.07 & 0.40 & 0.01 & 0.01 & 0.01 \\
spin-$\downarrow$& 0.00 & 0.00 & 0.03 & 0.23 & 0.08 & 0.01 & 0.01 \\
\end{tabular}
\end{ruledtabular}
\end{table}

\section{$5f$ orbital contribution near $E_F$ }

This array of figures provides the relative amounts of the stated
spin-orbital characters of states near the Fermi level. The bands on the
zone boundary X-U-R-S are flatter. More dispersion occurs along 
$\Gamma$-X and S-Y.

\begin{figure*}[!htbp]
\centerline{\includegraphics[width=2.0\columnwidth]{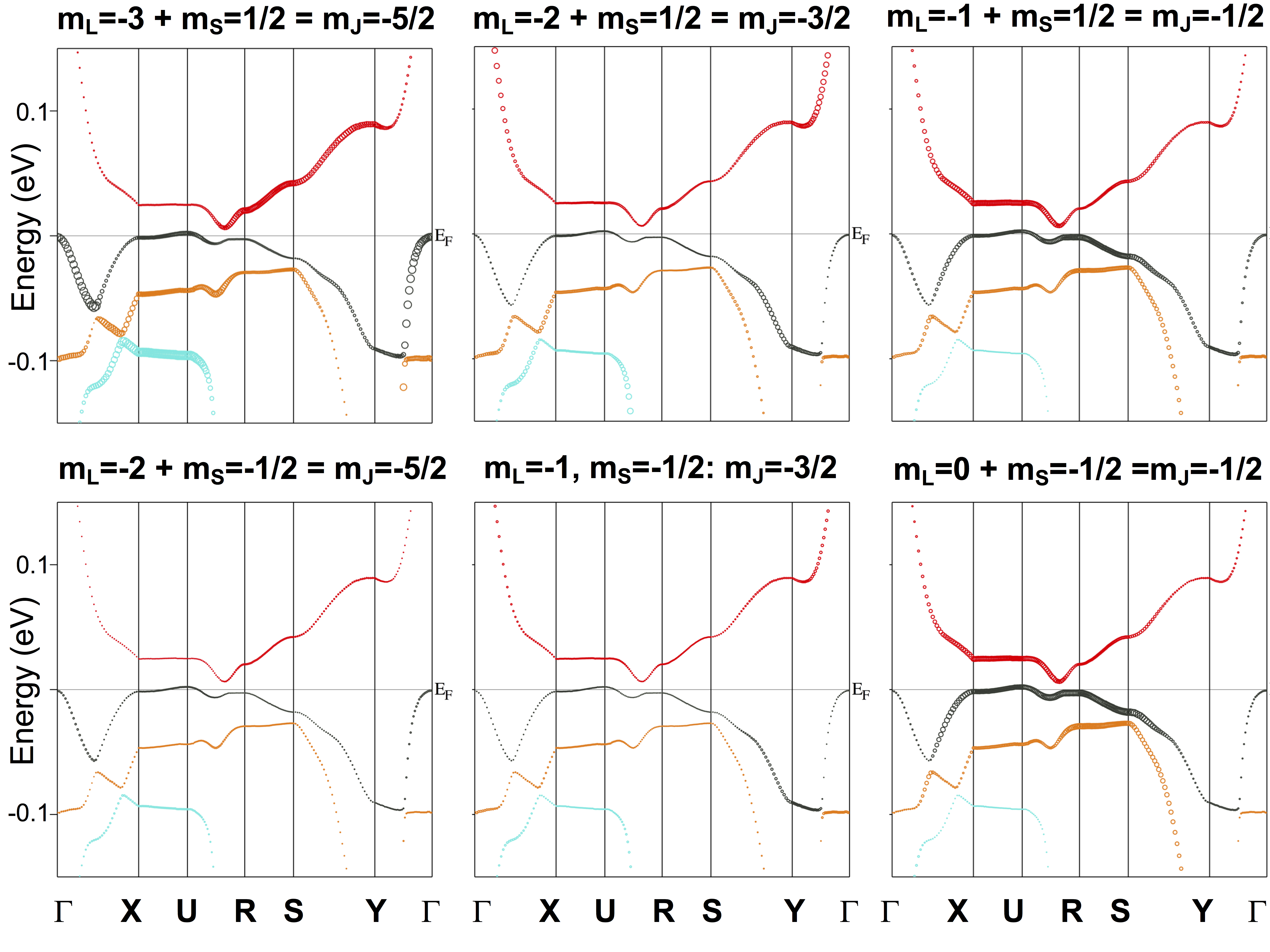}}
\caption{The $5f$ fat-band structure of non-magnetic UTe$_2$ from
LDA+U(OP) calculation. The circle size indicates the amount of
$j=\frac{5}{2}, m_j$ character in the bands, as indicated. The 
two contributions to the $j_z=-\frac{5}{2}$ character are quite
different; the differences in the other two cases is not so
pronounced.  Note that the full
energy range is only 300 meV.
}
\label{bnd2}
\end{figure*}

\end{document}